\documentclass[aps,10pt,longbibliography,pra,twocolumn,showpacs,amsmath,amssymb,floatfix,nofootinbib]{revtex4-1}

\expandafter\let\csname equation*\endcsname\relax
\expandafter\let\csname endequation*\endcsname\relax

\usepackage{lastpage}

\usepackage{fancyhdr}
\fancypagestyle{plain}{%
  \fancyhf{}%
}

\let\textcite\relax
\makeatletter
\DeclareRobustCommand{\MakeUppercase}[1]{{%
      \def\i{I}\def\j{J}%
      \def\reserved@a##1##2{\let##1##2\reserved@a}%
      \expandafter\reserved@a\@uclclist\reserved@b{\reserved@b\@gobble}%
      \protected@edef\reserved@a{\uppercase{#1}}%
      \reserved@a
   }}
\DeclareRobustCommand{\MakeLowercase}[1]{{%
      \def\reserved@a##1##2{\let##2##1\reserved@a}%
      \expandafter\reserved@a\@uclclist\reserved@b{\reserved@b\@gobble}%
      \protected@edef\reserved@a{\lowercase{#1}}%
      \reserved@a
   }}
\makeatother
\expandafter\let\csname ver@natbib.sty\endcsname\relax

\usepackage[style=alphabetic,backend=bibtex]{biblatex}
\DeclareFieldFormat{labelalpha}{\thefield{entrykey}}
\DeclareFieldFormat{extraalpha}{}

\usepackage{amsmath}
\usepackage[pdftex]{graphicx} 
\usepackage{dcolumn}  
\usepackage{bm}       
\usepackage[usenames,dvipsnames]{color}
\definecolor{URLCOL}{rgb}{0,0.17,0.43} 
\definecolor{LINKCOL}{rgb}{0.05,0.4,0} 
\definecolor{CITECOL}{rgb}{0.35,0,0.48} 
\definecolor{goodgreen}{RGB}{42,125,35} 
\usepackage{epstopdf}
\usepackage[pdftex,bookmarks,breaklinks,bookmarksopen,bookmarksnumbered,colorlinks,linkcolor=LINKCOL,linktocpage,citecolor=CITECOL,urlcolor=URLCOL,pdfpagemode=UseOutline,pdftex]{hyperref}

\usepackage{microtype}

\usepackage{booktabs}
\usepackage{verbatim}
\usepackage{comment}
\usepackage{multirow}

\definecolor{TITLECOL}{rgb}{0.1,0.2,0.7} 
\definecolor{PCOL}{rgb}{0.5,0.06,0.01} 
\definecolor{CHAPCOL}{rgb}{0,0.48,0} 
\definecolor{SECOL}{rgb}{0.1,0.2,0.7} 
\definecolor{CONTENTSCOL}{rgb}{0.1,0.2,0.7} 
\definecolor{SSECOL}{rgb}{0.25,0,0.48} 
\definecolor{SSSECOL}{rgb}{0.2,0.08,0.53} 
\definecolor{SHDCOL}{rgb}{0.4,0,0} 
\definecolor{ITMCOL}{rgb}{0.4,0,0} 
\definecolor{EXCOL}{rgb}{0,0.47,0.01} 
\definecolor{DEFCOL}{rgb}{0,0.42,0.01} 

\def\sec#1{\section{\textcolor{SECOL}{#1}}}
\def\ssec#1{\subsection{\textcolor{SSECOL}{#1}}}

\def\bea{\begin{eqnarray}}
\def\eea{\end{eqnarray}}
\def\ben{\begin{equation}}
\def\een{\end{equation}}
\def\bei{\begin{itemize}}
\def\eei{\end{itemize}}
\def\beit{\begin{itemize}}
\def\eit{\end{itemize}}
\def\benu{\begin{enumerate}}
\def\enu{\end{enumerate}}
\def\n{n}
\def\sss{\scriptscriptstyle\rm}

\def\1var{(\bx_1...\bx\N)}
\def\half{\frac{1}{2}}

\def\br{{\bf r}}

\def\bx{{x}}

\def\f{_{\rm frag}}
\def\d{_{\rm dis}}
\def\rl{_{\rm rel}}
\def\x{_{\sss X}}
\def\c{_{\sss C}}
\def\s{_{\sss S}}
\def\xc{_{\sss XC}}

\def\Hxc{_{\sss HXC}}

\def\N{_{\sss N}}
\def\H{_{\sss H}}

\def\hyb{^{\rm hyb}}
\def\HF{^{\rm HF}}

\def\GGA{^{\rm GGA}}

\def\PBE{^{\rm PBE}}

\def\TF{^{\rm TF}}

\def\ee{_{\rm ee}}

\def\dn{_\downarrow}

\def\sph_int{ {\int d^3 r}}

\def\intr{\int d^3r\,}
\def\intrp{\int d^3r'\,}

\def\deps{\Delta\epsilon_g}
\def\tdeps{\Delta\tilde\epsilon_g}

\def\a{\alpha}

\makeatletter 
\def\blx@maxline{77} 
\makeatother 

\begin{document}

\renewcommand{\bibliography}[1]{}

\thispagestyle{empty}

\title{The importance of being consistent.}

\author{Adam Wasserman}
\author{Jonathan Nafziger}
\author{Kaili Jiang}
\affiliation{Department of Physics and Astronomy, Purdue University, West Lafayette, IN 47907, USA}
\affiliation{Department of Chemistry, Purdue University, West Lafayette IN 47907, USA}

\author{Min-Cheol Kim}
\author{Eunji Sim}
\affiliation{Department of Chemistry and Institute of Nano-Bio Molecular Assemblies, Yonsei University, Seoul 120-749 Korea}

\author{Kieron Burke}
\affiliation{Department of Chemistry, University of California, Irvine, CA, 92697, USA}

\date{\today}

\begin{abstract}

We review the role of self-consistency in density functional theory.
We apply a recent analysis to both Kohn-Sham and
orbital-free DFT, as well as to Partition-DFT, which generalizes all aspects of standard DFT.  
In each case, the analysis distinguishes
between errors in approximate functionals versus errors in the self-consistent
density.  This yields insights into the origins of many errors in DFT
calculations, especially those often attributed to self-interaction or
delocalization error.  In many classes of problems, errors can be 
substantially reduced by using `better' densities.  We review the history
of these approaches, many of their applications, and give simple pedagogical
examples.

\end{abstract}


\maketitle


\def\T{\hat{T}}
\def\V{\hat{V}}
\def\eps{\epsilon}
\def\deps{\Delta\eps}
\def\dn{\Delta\n}
\def\Ts{\sqrt{n_1 n_2}}
\def\Uh{\frac{U}{4}\left(n_1^2 + n_2^2\right)}
\def\bUh{\left(n_1^2 + n_2^2\right)}  
\def\dndo{x}
\def\dndd{x^2}
\def\dndf{x^4}
\def\gm{\gamma}
 \newcommand{\sgn}{\operatorname{sgn}}
\tableofcontents

\def\a{\alpha}

\def\cm{{\bf\checkmark}}
\rm

\sec{Introduction}

Density functional theory(DFT) is used in more than 30,000 scientific papers per
year\cite{PGB15}.  
Most of these applications are routine, where the
calculation yields sufficiently accurate results as to provide insight
into some scientific or technological problem.  Most use the
Kohn-Sham (KS) scheme with one of a very small
set of popular functional approximations whose successes and failures
are well-documented.  For example, the standard approximations are
`known' to fail when there is substantial self-interaction or strong
correlation or localization in the system\cite{MCY08}.  These concepts 
are closely related to one another.

Partition-DFT (PDFT) is an exact generalization of DFT using
fragment densities as the basic variables\cite{CW07}.
Many difficulties of KS-DFT are overcome by PDFT.
PDFT can
deal with strong correlation; it allows for extremely chemical
interpretations of DFT calculations; it provides a direct route
to energy differences, not only total energies; and it is well suited for 
linear-scaling implementations and parallelization. 

Almost all DFT calculations employ a basic principle that was used
in the original Thomas-Fermi theory\cite{T27,F28}.  
When one constructs an approximation to the energy as a functional
of the density, one then uses it to find the density for that
system, by minimizing the approximate energy.   This is true
of the exact functional, and is used in almost all practical DFT
calculations with approximate functionals.   With such a choice,
basic theorems such as the Hellmann-Feynman theorem apply, allowing
easy calculation of forces, etc.

This principle appears so common-sensical that it is difficult
to question.  Surely you get the most accurate energy by minimization?
In fact, this is not always the case.
We examine the errors made by self-consistency 
and find that, in certain, well-defined, common situations, the
errors made in the density overwhelm those made in the evaluation
of the functional, and often these can be fixed with little additional
computational
effort.   

We also apply our enery-error analysis to PDFT, showing that several of
the key concepts in PDFT are, in fact, the same as those
involved in energy-error analysis, and how both can be employed
to understand the remaining errors in PDFT.

\begin{figure}[htb]
\begin{center}
\includegraphics[height=1.5in,width=3in]{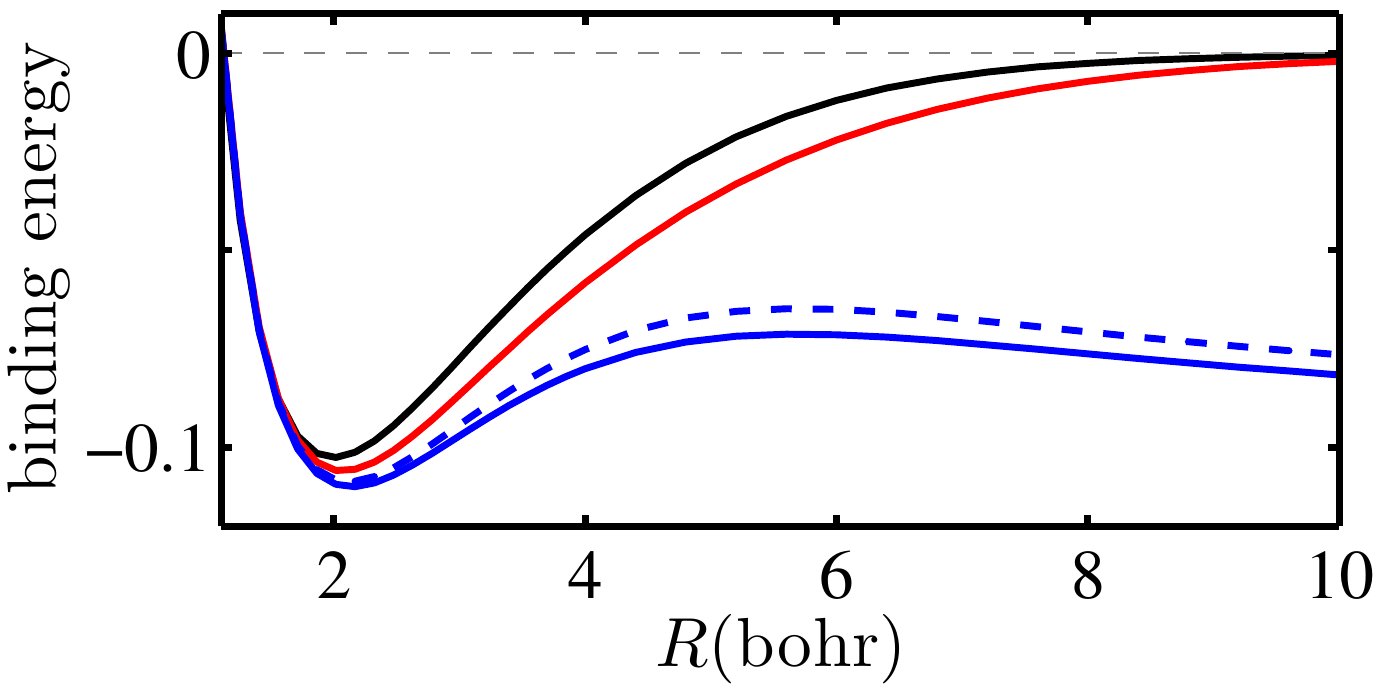}
\caption{Binding energy curves of H$_2^+$. Black is exact, blue is  self-consistent PBE, blue-dashed
is PBE on HF density, and red is approximate PDFT, Equation \ref{e:OA}. Energies are in eV.}
\label{f1}
\end{center}~
\vskip -1cm
\end{figure}

Throughout this article, we use the H$_{2}^{+}$ binding energy curve to
illustrate many concepts and approximations involved.  In {\bf Figure \ref{f1}},
the black line is the exact curve, given by a Hartree-Fock (HF) calculation,
while the blue line is for a standard DFT calculation\cite{PBE96}, showing the infamous failure as the bond is stretched\cite{MCY08}.  The blue-dashed line is from HF-DFT, 
meaning the DFT calculation on the HF density.  While this method cures
many problems with standard DFT, it has almost no effect here, because the
bond is symmetric.  On the other hand, a simple approximation within PDFT
(Section\ref{s:OA} within) yields a tremendous improvement over standard DFT.
The rest of this review explains how.

\sec{Background}
We restrict ourselves to non-relativistic systems within the Born-Oppenheimer approximation
with collinear magnetic fields\cite{ED11}.
DFT is concerned with efficient methods for finding
the ground-state energy and density of $N$ electrons whose Hamiltonian is
\ben
\label{hamiltonian}
\hat H = \hat T + \hat V\ee + \hat V,~~~
\hat V = \sum_{i=1}^N v(\br_i).
\een
The first of these is the kinetic energy operator, the second is the 
electron-electron repulsion, while the last is the one-body potential.
Only $N$ and $v(\br)$ change from one system to another, be they atoms,
molecules or solids.   We use atomic units throughout, unless otherwise
stated.

\ssec{Standard DFT}

\subsubsection{Pure DFT}

In 1964, Hohenberg and Kohn(HK)\cite{HK64} proved that, for a
given electron-electron interaction, there was
at most one $v(\br)$ that could give rise to the ground-state one-particle 
density $n(\br)$ of a system. If we write
\cite{L79,L83}
\ben
F[\n] = \min_{\Psi\to\n} \langle \Psi |\, \hat T + \hat V\ee\, |\Psi \rangle = T[n]+V\ee[n],
\label{F}
\een
where the minimization is over all normalized, antisymmetric $\Psi$ with
one-particle density $\n(\br)$, then
\ben
E = \min_\n \left\{ F[\n] + \int d^3r\, \n(\br) \,v(\br)\right\}.
\label{E}
\een
The Euler equation corresponding to the above minimization for fixed $N$ is
simply
\ben
\frac{\delta F[\n]}{\delta\n(\br)} = - v(\br).
\label{Euler}
\een
Armed with the exact $F[\n]$, the solution of this equation yields the exact ground-state density which,
when inserted back into $F[\n]$, yields the exact ground-state energy.

This theorem proved that the original, crude DFT of Thomas and Fermi\cite{T27,F28}
was an approximation to an exact approach.  Back then, they approximated
\ben
T[\n] \simeq T\TF[\n]=\frac{3(3\pi^2)^{2/3}}{10}\int d^3r\ \n^{5/3}(\br),
\label{TTF}
\een
and $V\ee$ with the Hartree energy, the classical self-repulsion of the
charge density
\ben
V\ee[n] \simeq U\H[n] = \frac{1}{2} \intr \intrp \frac{n(\br)\, n(\br')}{|\br -\br'|}.
\label{Hartree}
\een
Adding these together to approximate $F$ yields the iconic Thomas-Fermi(TF) theory,
and the Euler equation for an atom yields the TF density of atoms.
This approximation yields energies that are good to within about 10\%,
but since, e.g., all thermochemistry depends on very tiny differences
in electronic energies, TF theory is not accurate enough for chemical
or modern materials science applications.

\subsubsection{Kohn-Sham DFT} 


To increase accuracy and construct $F[\n]$, 
modern DFT calculations use the KS scheme that imagines a fictitious
set of non-interacting electrons with the same ground-state density as the real Hamiltonian\cite{KS65}.
These electrons satisfy the KS equations:
\ben
\left\{ -\half \nabla^2 + v\s(\br) \right\}\, \phi_i(\br) = \epsilon_i\, \phi_i(\br),
\label{KSeqns}
\een
where $v\s(\br)$ is defined as the unique potential such that 
$\n(\br)=\sum_{occ}\,|\phi_i(\br)|^2$.
To relate these to the interacting system, we write
\bea
F[\n]&=&T\s[\n]+E\Hxc[\n], \nonumber\\
T_s[\n] &= &\half \int d^3r\, \sum_{i=1}^N | \nabla \phi_i(\br)|^2,
\label{EHxc}
\eea
where $T\s$ is the non-interacting (or KS) kinetic energy, assuming the KS wavefunction (as is usually the case) is a single Slater determinant.
Here $E\Hxc=U\H+E\xc$ is the sum of the Hartree and exchange-correlation (XC)
energies and is {\em defined} by Equation \ref{EHxc}. Lastly, we differentiate Equation \ref{EHxc} with respect to the density, yielding
\ben
v\s(\br)
=v(\br) + v\Hxc(\br),
~~~v\Hxc(\br)=\frac{\delta E\Hxc}{\delta\n(\br)}.
\label{vsdef}
\een

\noindent This is the single most important result in DFT, as it closes the set of
KS equations\cite{KS65}.  
Since $U\H$ is known as an explicit density functional (Equation \ref{Hartree})
given any expression for $E\xc$ in terms of $\n(\br)$,
either approximate or exact, the KS equations can be solved self-consistently
to find $\n(\br)$ for a given $v(\br)$.  The self-consistency is simply
finding the minimum of an approximate $F$ determined from an approximate $E\xc$.
\begin{figure}[htb]
\begin{center}
\includegraphics[width=3in]{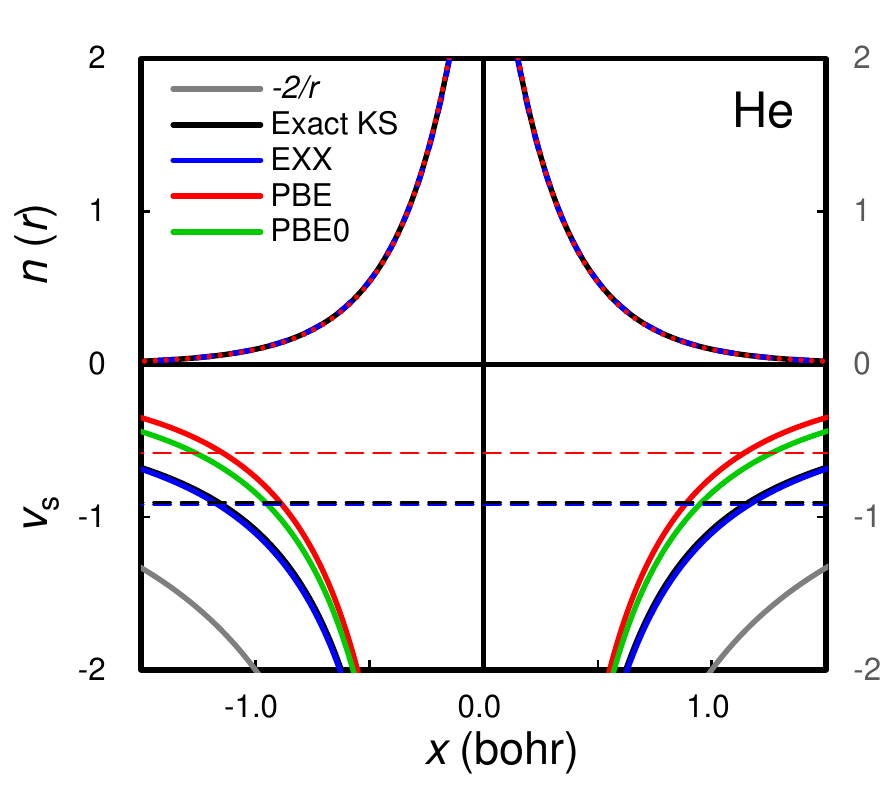}
\caption{Exact\cite{UG94} and approximate DFT densities and KS potentials
of the He atom, using PBE, PBE0 and exact exchange, in a.u.
The dashed horizontal lines indicate the eigenvalues of the 1s orbitals.}
\label{He}
\end{center}
~
\vskip -1cm
\end{figure}
In {\bf Figure \ref{He}}, we show the exact $v\s(\br)$ of the He atom, 
found by inverting Equation \ref{KSeqns}
after finding a highly accurate density by solving the Schr\"odinger equation\cite{UG94}.
Inserting two non-interacting KS electrons in the 1s orbital of $v\s(\br)$
yields the exact $\n(\br)$.  All practical KS-DFT calculations
approximate $v\s(\br)$.  The $1s$ HOMO is at precisely $-I$, where $I$ is the ionization
energy.  The energies and eigenvalues for both He and H$^-$, both exactly, given by quantum Monte-Carlo(QMC) densities, and approximately,
are given in Table I.

\begin{table*}
\caption{Energies for He and H$^-$ in Hartree.}
\label{tab1}
\begin{center}
\begin{tabular}{| c | c c c | c c | c c c |c c c|}
\hline
\multirow{2}{*}{atom} & \multicolumn{3}{c}{$E$}\vline & \multicolumn{2}{c}{$E_{\PBE}[\n]$} \vline& \multicolumn{3}{c}{$\Delta E_{\PBE} \times 1000$} \vline &\multicolumn{3}{c}{$\epsilon^{\rm HOMO}$} \vline\\
 & Exact & HF & PBE & $\n_{\rm HF}$ & $\n_{\rm QMC}$ & $\Delta E$ & $\Delta E_F$ & $\Delta E_D$ & exact & HF & PBE \\
\hline
He & -2.904 & -2.862 & -2.893 & -2.892 & -2.892 & 10.8 & 11.8 & -1.0 &-0.903 & -0.918 & -0.579 \\
H$^-$ & -0.528 & -0.488 & -0.538 & -0.521 & -0.527 & -10.4 & 1.0 & -11.4 &-0.028 &-0.046 & -0.000 \\
\hline
\end{tabular}
\end{center}
\end{table*}

Many forms of approximation\footnote{No approximate functional should be quite accurate.  It looks so calculating.}
 exist for $E\xc[\n]$,
the most popular being
the generalized gradient approximation
(GGA)\cite{P86,B88,LYP88,PCVJ92,PBE96}, and hybrids of GGA with exact exchange from a HF calculation\cite{B93,PEB96,AB99,HSE03},

\bea
E\xc\GGA&=&\int d^3r\ e\xc\GGA(\n(\br),|\nabla\n(\br)|), \nonumber\\
E\xc\hyb&= &a\, (E\x-E\x\GGA) + E\xc\GGA.
\eea

\noindent Here $a$ is the fixed mixing parameter, usually chosen between about 0.2 and
0.25 to optimize energetics for a large range of molecular dissociation
energies\cite{B93,PEB96}.
All practical calculations generalize the 
preceding formulas for arbitrary spin using spin-DFT \cite{BH72}.
The computational ease of DFT calculations relative to more accurate wavefunction
methods usually allows much larger systems to be calculated\footnote{In matters of density functional theory, reliability, not accuracy, is the vital thing.}, leading to DFT's immense popularity
today\cite{PGB15}.  
However, all these approximations fail in the paradigm case of
stretched H$_2$, the simplest example of a strongly correlated system\cite{B01}.

For just one particle,
we know the explicit functionals:
\ben
\label{vonWeisacker}
T\s = \int d^3r\,\frac{|\nabla\n|^2}{8\n},~~~
E\x=-U\H,~~~E\c=0,~~(N=1).
\een
None of the popular functionals satisfy these conditions for
all one-electron systems, and their errors are called self-interaction
errors (SIE).


In {\bf Figure \ref{He}}, $v\s\PBE(\br)$ is substantially above the exact curve,
and its HOMO level is several eV too high (Table I), but the almost constant shift in $v\s(\br)$ has little effect on $\n(\br)$ and therefore on $E$.
Note also that the HF potential is very close to
the true potential, and suffers none of the difficulties
of standard approximations.  But the hybrid functionals
have potentials that are essentially those of GGA with
$a$ times the HF potential mixed in, so their $\epsilon_i$
tend to have an error that is about a fraction $a$ smaller than
that of their GGA counterparts, i.e., still large, as
in the PBE0 curve of {\bf Figure \ref{He}}.
Many of these concepts are described more precisely these days
with the notion of delocalization error\cite{LZCM15,ZLZY15}.
These localization effects become more subtle in polarizable
solvent models\cite{DJ15}, and are especially important in Na-water
clusters\cite{SRR15}.

Later, we explain how such popular approximations for the energy can have
such `bad' potentials, yet yield such useful energetics.

\ssec{Partition DFT}

Most codes based on KS-DFT scale as $N^3$, with $N$ an appropriate measure of the size of the system. This is a very signiﬁcant improvement over correlated wavefunction-based methods, but still impractical for large systems. DFT-based Car-Parrinello optimizations, for example, are limited to systems of no more than a few thousand atoms. In response to this challenge, linear-scaling schemes have been developed \cite{G99}. Some of these take advantage of the nearsightedness of electronic properties \cite{SK66,Y91}. 
 Other schemes break the system into fragments that are small enough for rapid computation, and then build the properties of the whole system in a way which preserves order-$N$ scaling \cite{NG04,FK07}. Since the unfavorable scaling of KS-DFT arises primarily from the use of KS orbitals, orbital-free schemes have also been developed that perform direct minimization of the energy functional and scale linearly with $N$ \cite{WC00,HC09}. 
The quasicontinuum-DFT approach (QCDFT) \cite{PZHC08}, combining the coarse-graining idea of multiscale methods \cite{CLK05} with the coupling strategy of QM/MM (Quantum-Mechanics / Molecular-Mechanics) \cite{SHFM96,GT02,FG05}, allows for the simulation of multimillion atoms via orbital-free DFT embedding. Explicit treatment of a few million atoms has been demonstrated via linear-scaling orbital-free DFT algorithms \cite{HC09,CJZ+16}.  
 These, however, rely on approximations to the non-interacting kinetic energy functional $T_s[n]$, which are neither suf{f}iciently accurate nor general. 


PDFT\cite{CW07,EBCW10} is an exact reformulation of DFT with the potential
to overcome both problems of scaling with system size and problems related
to errors made by the approximate XC functionals. PDFT was developed initially
to strenghten the
foundations of chemical reactivity theory \cite{CW07, CW03}.
Its structure belongs to the family of density-based embedding methods
that were developed starting in the early 1970's to improve the efficiency of
 electronic-structure calculations via fragmentation. PDFT generalizes all aspects
of both pure and KS-DFT with new variables that have an extremely chemical interpretation, 
while also providing all the computational advantages of quantum embedding methods.
Because excellent, comprehensive reviews on embedding have appeared recently \cite{JN14,KSGP15,WSZ15},
we list only a few highlights relevant to this review.


\subsubsection{A few quantum embedding highlights}

{\hspace{1cm}} \\ \indent {\em 1970}'s: Based on the assumption that the density of rare-gas dimers can be well approximated by the sum of their isolated-atom densities, the first non self-consistent embedding calculations of the binding-energy curves of rare-gas dimers were performed by Gordon and Kim(GK method). \cite{GK72}. 

{\em 1980}'s: Corrections were added to the non-self consistent GK calculations to account for self-interaction errors \cite{WP81} and to include induction effects and dispersion forces \cite{H84}. The first self-consistent versions of the GK model were also proposed \cite{SS86}.

{\em 1990}'s: Subsystem-DFT (S-DFT) \cite{C91} and frozen-density embedding (FDE) \cite{WW93} were developed. 
FDE was initially not completely self-consistent, but was later made self-consistent via freeze-and-thaw cycles,
making it equivalent to S-DFT. 
The self-consistent atomic deformation theory (SCAD) is a version of S-DFT requiring the
fragment densities to be written as atomic densities \cite{BM93}. 
Other methods treat different fragments with different levels of theory, allowing for critical fragments
of a larger calculation to be treated with higher accuracy
(usually referred to as embedding-DFT). In all cases,
the main equations are the KS equations with constrained electron density (KSCED) \cite{WW96}. 

{\em 2000}'s:  Many developments took place, mostly of a technical nature \cite{HC08}:  FDE was
applied with a plane-wave basis and both local and non-local pseudopotentials \cite{TB00}; 
the idea of buffer fragments was introduced\cite{CW04}; 
FDE was extended to time-dependent DFT (TDDFT) \cite{Wes04,NLBW05,Neu07} 
 and to work in combination with con{f}iguration-interaction methods \cite{KGWC02}. 
 In parallel, significant advances were made for computational sampling procedures in QM/MM \cite{KHW09}.

{\em 2010}'s:  New methods can now calculate $\delta T_{\s}^{\rm nad} / \delta \n(r)$ for covalent bonds \cite{FJNV10},
or bypass the need for inversions altogether via exact density embedding \cite{MSGM12}; 
FDE develops to study charge-transfer reactions \cite{PN11}, 
calculate charge-transfer excitation energies and diabatic couplings \cite{PVVN13}, 
and include van der Waals interactions \cite{KEP14}. 
Much is now known about the performance of approximate self-consistent S-DFT \cite{SKMV15}.
Sources of error in WFT-in-DFT embedding was investigated \cite{GBMMI14}. 
 

\subsubsection{PDFT in a nutshell}

Although PDFT has been extended to the time-dependent case \cite{MJW13,MW14,MW15}, we
focus here on the ground-state theory, where the
goal is to calculate $E$ and $\n(\br)$ of a molecule via fragment calculations.
The user chooses how the nuclei are assigned into fragments by dividing the one-body potential.
For simplicity, we give formulas for just two fragments, but there can be as many as desired. Here
\ben
v(\br)=v_1(\br)+v_2(\br).
\label{e:fragment_v}
\een
The choice of $\{v_1, v_2\}$, together with $N$, unambiguously determine a unique, global {\em partition potential} $v_p(\br)$ and a unique set of fragment densities \cite{CW06}:  ${\bf \n}=(\n_1(\br),\n_2(\br))$. Each resulting $n_\a(\br)$ is the ground-ensemble density of
$N_\a$ electrons in $v_\a(\br) + v_p(\br)$, with $N_1+N_2=N$.
At self-consistency, $v_p(\br)$ is global (independent of $\a$), and
\ben
\n_1(\br)+\n_2(\br) = n(\br).
\label{e:fragment_n}
\een
We omit here spin indices for notational simplicity (but see \cite{MW13, NW14}).
The self-consistent equations that are solved to find $\bf \n$
and the partition potential $v_p(\br)$ follow from the Euler equation of
a constrained minimization. The quantity being minimized is not
the total energy of the molecule, as in standard FDE and S-DFT, but rather
\ben
E\f [{\bf \n}]=E_1[\n_1]+E_2[\n_2],
\label{e:Ef_def}
\een
the sum of the fragment energies, where $E_\a[\n]$ is the ground-state
energy functional for potential $v_\a(\br)$.
If $N_1$ is not an integer, then write $N_1=M+\nu$, $0 \leq \nu < 1$, and
\cite{PPLB82}:
\bea
E_1[n]&=&(1-\nu) E_1[n_M]+\nu E_1[n_{M+1}], \nonumber\\
n_1(\br)&=&(1-\nu) n_M(\br)+\nu n_{M+1}(\br).
\label{e:PPLB}
\eea
Thus, only integer calculations need be performed,
but $E\f[{\bf \n}]$ is minimized with respect
to $\nu$ as well, so $N_1$ need not
be an integer.

The formal constraints under which $E\f[{\bf \n}]$ is minimized
are Equation \ref{e:fragment_n} and the number constraint $N=N_1+N_2$.
The partition potential $v_p(\br)$ and the chemical potential $\mu$ can
be seen as the Lagrange multipliers guaranteeing these constraints.
Writing $E_{1,2}[n]$ in terms of KS quantities, this constrained minimization leads to the KS-PDFT
equations\cite{EBCW10} which, for a given approximation to the $\rm XC$ functional,
{\em exactly} reproduce the results of the corresponding KS calculation for the
entire system (using the same $\rm XC$ functional).
At the minimum, $E\f [{\bf \n}]$ differs from the
true energy by the {\em partition energy} $E_p[{\bf \n}]$, whose functional derivative evaluated at any minimizing $n_\a(\br)$
is the partition potential:

\ben
E_p[{\bf \n}]\equiv E[\n] -E\f[{\bf \n}], ~~ v_p(\br)=\frac{\delta E_p}{\delta \n_1(\br)}
=\frac{\delta E_p}{\delta \n_2(\br)}.
\label{e:Ep_def}
\een
As in S-DFT, the partition energy is divided into the non-additive Kohn-Sham components:
\ben
E_p[{\bf \n}]=F^{\rm nad}[{\bf \n}]+V^{\rm nad}[{\bf \n}]
\label{e:Ep_split}
\een
where $F^{\rm nad}[{\bf \n}]=F[n_1+n_2]-F[n_1]-F[\n_2]$. In Equation \ref{e:Ep_split}, $V^{\rm nad}$ includes both the non-additive electron-nuclear and nuclear-nuclear interactions.
The calculation requires either an explicit density-functional approximation for $T_s^{\rm nad}[{\bf \n}]$, as in Ref.\cite{WEW98},
or (computationally expensive) inversions, as in Ref.\cite{GAMM10}.
If one only minimizes $E[n]$, this non-additive term
may be made to vanish by requiring that orbitals from different fragments are orthogonal to
each other\cite{MSGM12}. This, however, requires a molecular KS calculation ahead of time.

\begin{figure*}[htb]
\begin{center}
\includegraphics[height=1.5in,width=5in]{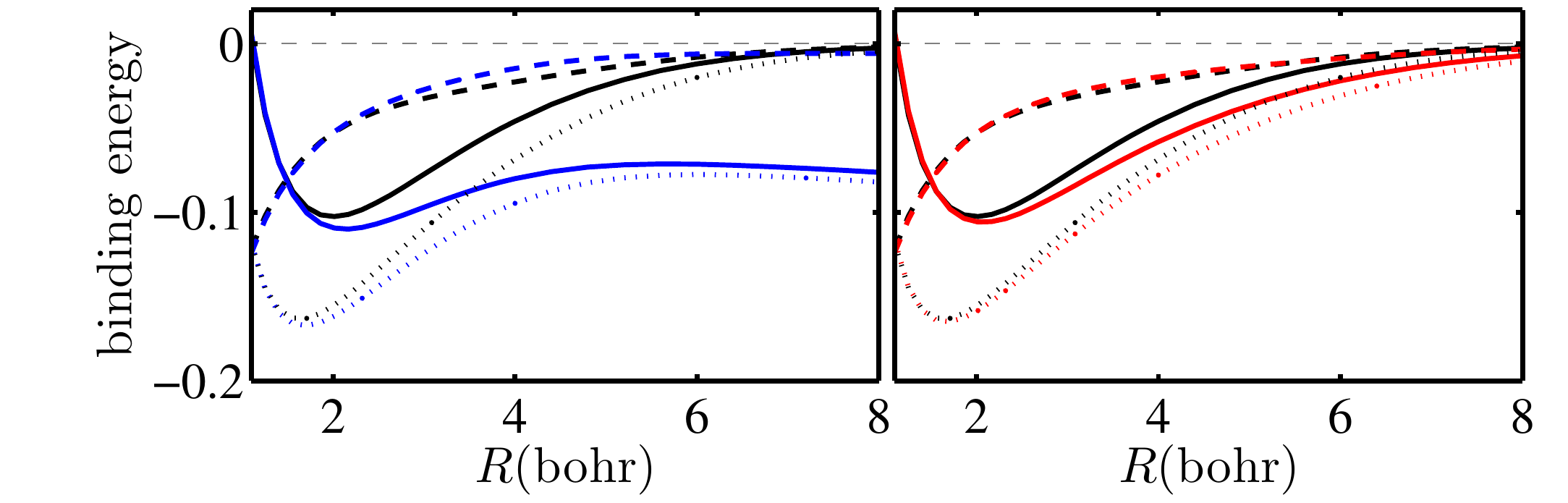}
\caption{Partitioned energy contributions to binding curve of H$_2^+$.  
The dissociation curve ($E\d$, solid) is the difference between the partition energy ($E_p$, dotted, $1/R$ included) and the relaxation energy ($E\rl$, dashed), Equation \ref{e:E_p_decomp}. Black are exact, blue are PBE, red are the overlap approximation.}
\label{f3}
\end{center}~
\vskip -1cm
\end{figure*}

Why minimize $E\f[{\bf \n}]$ (Equation \ref{e:Ef_def}) rather than
the total energy directly? With the constraint of Equation \ref{e:fragment_n}, the answer is clear:
When $n(\br)$ is the true ground-state density,  the HK theorem guarantees that
we have {\em also} minimized $E[n]$, and produced ``chemically meaningful" fragments\cite{RP86}.
The total work done in deforming the isolated fragment densities to produce the PDFT fragment densities is
the relaxation energy $E\rl$,
\ben
E\rl=E\f^{(\infty)}-E\f,
\label{e:Erel}
\een
where $E\f^{(\infty)}$ is the sum of the fragment energies when the fragments are infinitely separated from each other. (In the original GK model \cite{GK72},  $E_{\rm rel}=0$.) The true dissociation energy of the system, $E\d =E-E\f^{(\infty)}$ is related to the partition energy, Equation \ref{e:Ep_def}, by:
\ben
E_p=E\rl+E\d.
\label{e:E_p_decomp}
\een
In {\bf Figure \ref{f3}}, we show the exact contributions and their PBE counterparts.  
Both $E\rl$ and $E_p$ contribute substantially at equilibrium.
Clearly, the failure of PBE is primarily in $E_p$.

The partition trick is thus analogous to the KS trick:  The former maps the system
into isolated fragments, while the latter maps the system to non-interacting
electrons\cite{N15}.
In KS-DFT, the self-consistent density from solution of the KS equations is also that which minimizes $E[n]$. The KS ``density constraint"  guarantees this, by construction.
Furthermore, $v_p(\br)$ in PDFT, like $v\Hxc(\br)$ in KS-DFT, is a global potential that
is added to $v(\br)$ to make the auxiliary system. It is {\em unique} for a choice
of partitioning, as follows from the minimization of $E\f[{\bf \n}]$ \cite{CW06}.
In this analogy, PDFT is to subsystem-DFT like KS-DFT
is to Hartree-Fock theory.

\subsubsection{In practice: Converging to self-consistency}

For each fragment $\a$, two KS-like equations are solved simultaneously:
\ben
\left\{-\half\nabla^2+v_\a^{\rm eff}[n_\a^{\pm}](\br)
+v_p(\br)\rm \right\}\phi_{i,\a}^{\pm}(\br)=\epsilon^\pm_{i,\a}\phi^{\pm}_{i,\a}(\br),
\label{e:KS-like}
\een
where the effective potential is just the usual KS form, Equation \ref{vsdef}, and $\pm$ denotes evaluation for $M$ and $M+1$ electrons.
The various partition potentials 
generally differ until a self-consistent solution is reached.
For a given set of trial fragment densities, define
\ben
v_{p,\a}^\pm(\br)=\delta E_p[{\bf \n}] / \delta n^\pm_\a(\br).
\een
We construct a weighted average partition potential over all fragments and particle numbers:
\ben
v_p(\br)=\int d^3r'\, \sum_{\a=1}^2\, \sum_{\lambda=\pm}\, v_{p,\a}^\lambda(\br')\, Q^\lambda_{\a}(\br',\br).
\label{e:vp_average}
\een
The $Q$-functions provide the bridge between PDFT and S-DFT calculations\cite{NW14},
and are approximated in practice as\cite{MW13}:
\ben
Q^\lambda_{\a}(\br',\br)=\frac{\delta n^\lambda_{\a}(\br')}{\delta n\f(\br)}
\approx\frac{n_{\a}^\lambda(\br)}{n\f(\br)}\delta(\br-\br')~~.
\label{e:local_Q}
\een
In Equation \ref{e:local_Q}, $n\f(\br)$ is the sum of
{\em trial} fragment densities at intermediate iterations, equal to the
correct molecular density only at convergence.
When the exact partition energy is used, either via iterative inversions\cite{NWW11} or through use
of the exact $T\s[\n]$, any approximate $Q$-functions such as Equation \ref{e:local_Q}, satisfying the sum-rule:
\ben
\sum_{\a=1}^{2}\sum_{\lambda=\pm} Q^\lambda_{\a}(\br',\br)=\delta(\br-\br')~~,
\een
will lead to the optimal $v_p(\br)$. However, it remains to
be investigated how the solutions depend on the choice of $Q$-functions when
approximations for $E_p[{\bf \n}]$ are employed.

\subsubsection{Overlap approximation}
\label{s:OA}

For a given XC approximation, the exactly corresponding $E_p[{\bf \n}]$ reproduces
the results of a molecular KS calculation, including all of the errors of the
underlying XC functional.  Carefully constructed
approximations to $E_p[{\bf \n}]$ have the potential to eliminate some of these errors, because $E_p$ can depend on individual fragment densities.  
An {\em overlap approximation} (OA) significantly reduces the delocalization and static-correlation errors of
semi-local XC functionals. The OA approximates the $E\Hxc^{\rm nad}$ contribution of Equation \ref{e:Ep_split} as:
\ben
\tilde{E}\Hxc^{\rm nad}[{\bf \n}]
=U\H^{\rm nad}[{\bf \n}]+S[{\bf \n}] E\xc^{\rm nad}[{\bf \n}]+(1-S[{\bf \n}])\Delta U\H^{\rm nad}[{\bf \n}]~~,
\label{e:OA}
\een
where $S[{\bf \n}]$ is an appropriate measure of the
spatial overlap between fragments and $\Delta U\H^{\rm nad}[{\bf \n}]$ is a correction to the non-additive Hartree designed to be used with semi-local XC-functionals\cite{NW15}.
The right panel of {\bf Figure \ref{f3}} for H$_2^+$ shows how the OA, when used with PBE for
the fragments, greatly improves the dissociation curve, getting the
stretched limit correct.  It even improves the fragment energies.
Self-consistency within PDFT works well.

\sec{A theory of inconsistency}
\label{s:PI}

In almost all DFT calculations, we use the HK theorems in two
ways simultaneously.  We make some approximation to an
energy, as a functional of the density, {\em and} we use the Euler
equation (or equivalently the KS equations or the partition
equations) to find the density that minimizes that energy functional.  Since
such equations are often solved by an iterative process, the
solution is usually called self-consistent.

But here we will explore how such a procedure might not always yield
the most accurate result for a given approximation.  Our standard approximations have been designed to yield reasonably accurate energetics for the Coulombic systems that nature has given us, but not accurate functional derivatives ({\bf Figure \ref{He}}).  Usually, the kinds
of inaccuracies in these derivatives are not very important but
as we show, sometimes they are very important.  Thus we consider
performing DFT calculations in which the density is {\em not} the
self-consistent solution with a given approximate energy, i.e., density
and energy are approximated separately.  For very
good, well-understood, reasons, such inconsistent density
functional calculations (IDFC's) can sometimes yield much
more accurate energies than self-consistent DFT calculations.

Our basic tool in analyzing such IDFC's will be the {\em energy-error
analysis}.
In practical DFT calculations,
$F[\n]$ is approximated,
call it $\tilde F[\n]$.  The minimizing
density $\tilde\n(\br)$ is therefore also approximate.
The energy-error is
\bea
\Delta E&=&\tilde E[\tilde \n] - E[\n] = \Delta E_F + \Delta E_D,\nonumber\\
\Delta E_F&=&\tilde E[\n] - E[\n],~~ \Delta E_D = \tilde E[\tilde \n] - \tilde E[\n]
\label{DE}
\eea
where $\Delta E_F$ is called the functional (or {\em energy-driven})
error, $\Delta E_D$ is the density-driven error, and they sum
to the total energy-error.   This single line of arithmetic is a powerful tool for analyzing errors in approximate DFT calculations.

Since the energy-error of {\em any} approximate self-consistent DFT calculation
can be decomposed in this manner, we choose the following
classification scheme.  We call a DFT calculation {\em normal}
if, for the energy of interest, $|\Delta E_F |>> |\Delta E_D |$.  
The vast majority of present DFT calculations meet this criterion,
which is why we call this normal.  On the other hand, if 
$|\Delta E_D| \approx |\Delta E_F|$ or larger, the calculation is {\em abnormal}.  Then the error in the energy of interest is typically substantially reduced if a more
accurate density than the minimizer of $\tilde F$ can be found.

Note that classifying a calculation as abnormal depends on
(a) the approximation used, (b) the system, and (c) the energy
of interest.   In applications of ground-state 
DFT, overwhelmingly the quantity of interest is {\em not} the density,
but rather the ground-state energy of the electrons.  This includes
all geometries, bond energies, vibrational frequencies, transition
state barriers, ionization energies and electron affinities, and
even polarizabilities, which can be deduced from changes in the
energy as a weak field is applied.

\ssec{Toy model}

To illustrate the general idea, consider a problem where we wish to find
a function
\ben
e_y=\min_x \left( f(x) - y\, x\right)
\een
where $f$ is an exact function, while $\tilde f$ is some approximation 
to it.   For example, choose $f(x)=\beta x^2$, where $f(x)$ is exact when $\beta = 1$.
Thus $\tilde f$ is a good approximation if $\beta = 0.9$, being within 10\% of $f$ for
all $x$.  In general, we can differentiate to find the minimizer:
\bea
f'(x_m)&=&y,~~~~~~x_m=[f']^{-1}(y),\nonumber\\
e_y&=&\left(f(x)-x\, f'(x)\right)\bigg|_{x_m(y)}
\eea
In our specific case, $x_m = y/(2\beta)$ and
$e_y=-y^2/(4\beta)$. Then the error in $\tilde e_y$ is just Equation \ref{DE}:
\bea
\tilde{e}_y - e_y&=&\Delta e_y = \frac{1-\beta}{\beta}\, e_y,\nonumber\\
\Delta e_F&=&(1-\beta)\, e_y,~~ \Delta e_D= \frac{(\beta-1)^2}{\beta} e_y.
\eea
Since here $\beta=0.9$, $\Delta e_F$ is slightly smaller than
$\Delta e_y$, while $\Delta e_D$ is much smaller.  This is a perfectly normal system.

\begin{figure*}[htb]
\begin{center}
\includegraphics[height=1.5in,width=5in]{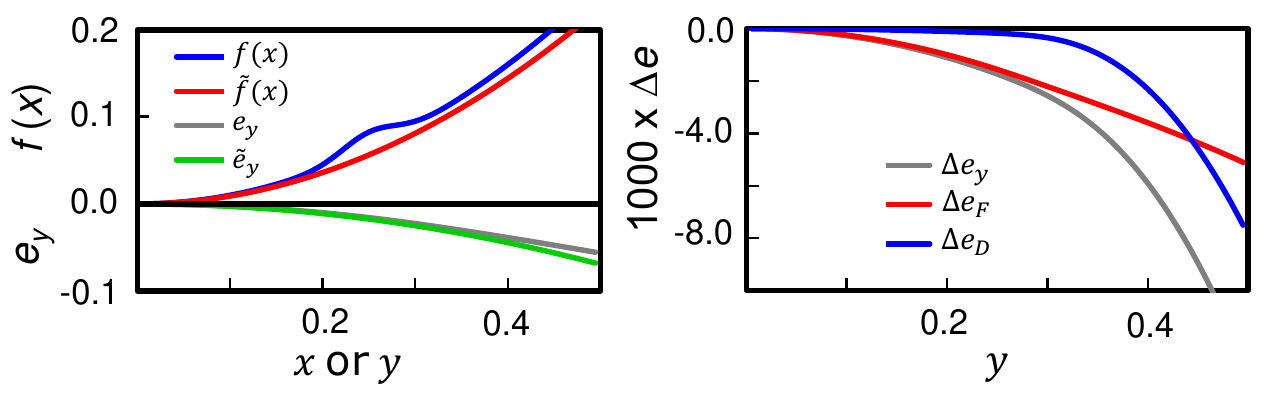}
\caption{(Left) Exact function and its approximation (versus $x$) above 0,
and $e_m$ and its approximation (versus $y$) below 0. (Right) Error at minimum and its decomposition for the toy
model.}
\label{tmod}
\end{center}
~
\vskip -1cm
\end{figure*}

But watch what happens when we add a small Gaussian, $a \exp(-[(x-b)/c]^2/2)$
to $f(x)$, where $a$ is 0.02,
$b$ is 0.25, and $c=0.03$.
 This has a relatively small effect on $\tilde f$, and
even on $\tilde e_y$, as shown in {\bf Figure \ref{tmod}}. However, consider the right panel in {\bf Figure \ref{tmod}}, which shows the total error and its decomposition
as a function of $y$.  For $y \leq 0.3$, the system is normal, and $|\Delta e_D | << |\Delta e_y|$.  But as
we approach $y=0.4$, $|\Delta e_D |$ grows much more rapidly than
$|\Delta e_F |$, and even becomes larger than it after $y=0.4$.

How has this happened?  The feature we added is not large, but it does vary
rapidly.  Thus $\tilde f \approx f$ everywhere, but $\tilde f'$ is {\em not}
close to $f'$.  This causes a large error in $x_m$ which produces a large error in $\tilde e(y)$, whose
origin is quite different from the normal case.  A careful expansion about the
exact and approximate minima yields:
\ben
\Delta e_D / \Delta e_F = - (\Delta f'_m)^2 / (f''_m\Delta f_m)
\label{DFfrac}
\een
where $\Delta f = \tilde f - f$.  In the normal case, $\Delta f,\Delta f',\Delta f''$ are all comparable in size so Equation \ref{DFfrac} is small. 
we see that $\Delta e_D$ is much smaller than $\Delta e_F$.  
But if $|\Delta f'_m| \geq \sqrt{f''_m\, |\Delta f_m|}$, then the calculation is abnormal, and $|\Delta e_D |$ dominates $|\Delta e|$. In the specific case we just calculated, we find $y=2 x_m$, so when
$x_m = 0.1$, the center of the Gaussian, $y$ is 0.2, and the system becomes abnormal.

\ssec{Extension to functionals}

We can apply everything from our toy problem to the minimization
of approximate functionals.  The density of any approximate
DFT calculation satisfies:
\ben
\frac{\delta \tilde F}{\delta \n(\br)}\Bigg|_{\tilde\n} = - v(\br) ~~~ {\rm or} ~~~ \tilde \n(\br)= \left\{\frac{\Delta \tilde F}{\delta\n}\right\}^{-1}[-v](\br).
\label{fdev}
\een
Just as in the toy, even if $\tilde F \approx F$ near $\n(\br)$,
rapid changes in $F$ with $\n(\br)$ that are not in $\tilde F$ can produce
unusually poor densities, leading to density-driven errors.  
It is well-known\cite{PPLB82} that total energies have derivative discontinuities
at integer values of $N$ and that these are also present in the exact
$F[\n]$, while standard approximations that are explicit density functionals
(such as TF and GGA) produce smooth functions of $N$.  Thus, whenever
such discontinuities are important, we should watch out for density-driven
errors.

Taking another functional derivative of Equation \ref{fdev} with respect to $v(\br)$ yields
the density change in response to a perturbation:
\ben
\delta \tilde\n(\br) = \int d^3r'\, \tilde\chi[\n](\br,\br')\, \delta v(\br')
\een
where $\tilde\chi$ is the (static) density-density response function, and is
the inverse of $\delta\tilde F/\delta\n(\br)\delta\n(\br')$.  
By analogy with Equation \ref{DFfrac}, the ratio $\Delta E_D/\Delta E_F$
is proportional to $\tilde\chi$.  An unsually large response function suggests
a significant density-driven error.

Pure DFT calculations, at least those with approximations dominated by the TF approximation, are always abnormal, i.e., the error is always density-driven.  On the other hand, most modern self-consistent KS-DFT calculations have excellent densities and are normal.  However, in a variety of well-known situations, the density-driven error with standard approximations
(GGA and hybrids) becomes unusually large, and dominates the error in
the calculation.  Such errors can all be greatly reduced by using a more
accurate density.  Finally, in the last section, we show how PDFT errors
can be understood with this analysis.

In fact, we can do a simple case exactly\cite{B07}.
Consider same-spin non-interacting
fermions in a flat box in one dimension, the simple
problem usualy done first in any quantum text book.  The potential energy
is zero everywhere, and the total energy is all kinetic.  For one particle
in a box of length 1,
$T=\pi^2/2$ exactly.  On the other hand, the TF approximation for such
a problem is:
\ben
T\TF[\n] = \frac{\pi^2}{6} \int dx\, \n^3(x).
\een
Minmizing in the box yields a constant density, $\n=1$.  Thus $T\TF=\pi^2/6$,
being too small by a disastrous factor of 3.
However, if we insert $n(x)=\sin^2(x)/2$ into $T\TF[\n]$, we find a much
better answer, $T\TF[\n]= 5 \pi^2/12$, i.e., we are now only too small by 1/6.
Thus the TF functional is far more accurate on the exact density than the
self-consistent one.  In terms of our energy-error analysis, we find
\ben
\Delta E_F = \frac{1}{4}\Delta E,~~~~\Delta E_D = \frac{3}{4}\Delta E,
\een
i.e., the density-driven error is three times larger than the functional-error.
These features remain true for all values of $N$.  Although TF theory becomes relatively exact here as $N\to\infty$, the density-driven error is always three times larger than the
functional-error, and dominates the energy-error.  This calculation is always
abnormal.

\sec{Pure DFT}

We begin with simple examples that can be easily done with Mathematica.
Consider the Bohr atom, which is an atom in which the electron-electron
repulsion has been turned off\cite{HL95}.  The orbitals are purely hydrogenic,
and the energies are those of a sum of the lowest hydrogenic levels.
Solving the Euler equation yields the TF density for this problem:
\ben
\n\TF(r)=\frac{4Z}{\pi^2 r^3_c}\, \left(\frac{r_c}{r}-1\right)^{3/2} \Theta (r_c-r),
\label{nTFbohr}
\een
where $r_c \!=\!  (18/Z)^{1/3}$, and $\Theta$ is the Heaviside step function.
The TF energy is just $-Z^2 (3N/2)^{1/3}$, where $Z$ is the
nuclear charge and $N$ the number of occupied orbitals.  
For $Z = 1$, this yields a ground-state energy of $-1.144$, which is more than
double the exact answer of $-1/2$. 
On the other hand, evaluating the TF kinetic energy on the hydrogen atom
density yields:
\ben
E\TF[\n]=\frac{81 (3 \pi)^{2/3}}{1250}-1 \approx -0.711,
\een
which is (only) a 40\% overestimate in magnitude, and the calculation is abnormal. TF errors are similar for real atoms. In radon, $\Delta E^{\TF} \simeq -3400$, and the relative energy-error vanishes as $Z \rightarrow \infty$\cite{LS73}. But $\Delta E_F \simeq -620$, so most of the energy-error is due to the density-error. There is no reason to think that this behavior would be any different in molecular calculations, or calculations of insulating solids.   It might change for simple metals with a pseudopotential, where the (pseudo)density is closer to slowly-varying.

Standard approaches to orbital-free DFT that are dominated
by local and semilocal approximations are likely to have errors
dominated by the density.  Calculations that test kinetic
energy functionals on the exact KS density rather than self-consistently
will typically have much smaller errors than self-consistent calculations.
Furthermore almost all semilocal approximations fail to converge in self-consistent calculations. \cite{XC15} 
The focus should be on improving the functional
derivative rather than the energy itself,  and the measure of improvement
should be the reduction of the density-driven error.

An entirely new method of finding the kinetic-energy
functional has recently appeared, using machine-learning to learn from
solved cases\cite{SRHM12,LSPH15,VSLR15}.  But its functional derivatives are so poor that they
are totally unusable for finding a self-consistent solution. Several techniques have been developed which constrain a minimization to stay
on the manifold of densities on which the machine-learned functional is
accurate\cite{SRHB13,SMBM13}.  These lead to algorithms that produce accurate densities, although
the density-driven error is up to 10 times greater than the functional
error, and the solutions also are slightly dependent on the starting point.
This has led to attempts to map the density-potential functional directly,
bypassing the need for an accurate functional derivative\cite{BLBM16}.

\sec{KS-DFT}
\label{ksdft}

We next apply the principles of inconsistency to KS-DFT. 
The KS scheme is simply an elaborate way to minimize
an approximation to $F$, given by Equation \ref{EHxc}.  All the same principles apply to $\tilde E\xc[\n]$ as we have already discussed about $\tilde F$.
Because GGA and hybrids use continuous
explicit density functional approximations, they miss the derivative
discontinuity, which shows up in the XC functional.   Thus their derivatives
are highly inaccurate, as in {\bf Figure \ref{He}}.  The KS potential of these
approximations is too shallow by several eV, yielding poor orbital energies, but the potentials are almost perfect constant shifts relative to the
exact potentials, at least within most of the atom or molecule.  Such
a shift has no effect on the shape of the orbitals, and therefore
on the density.  In fact, most KS-DFT calculations have excellent densities so even for cases with poor results, their errors are functional-driven, not density-driven\cite{TS66}. 
For the He atom of {\bf Figure \ref{He}}, $\Delta E_D$ is -9\% of 
$\Delta E$ in PBE. 
The functional error dominates 
and
the error in PBE worsens if the
exact density is used.  
Thus, all KS-DFT calculations with the standard functionals have poor-looking
KS potentials.  In a certain subset of cases, these poor quality potentials 
will lead to sufficiently poor self-consistent densities that density-driven
errors become significant.  Such calculations are abnormal and, if a more
accurate density is available, the error reduces significantly.

Is there any way to know, a priori, if a given approximate DFT calculation
is likely to suffer from a density-driven error?  There is. The KS response function is 
\ben
\tilde\chi\s(\br,\br')=\sum_{i,j} (f_i-f_j) 
\frac{\tilde\phi_i^*(\br)\tilde\phi_j^*(\br)\tilde\phi_i(\br')\tilde\phi_j(\br')}{\tilde\epsilon_i-\tilde\epsilon_j + i0_+} ~,
\een
where $f_i$ is the KS orbital
occupation factor\cite{DG90}.  The smallest denominator is $\tdeps$,
the HOMO-LUMO gap.
Normally, the difference between the exact and approximate $v\s(\br)$
is small, ignoring any constant shift. If
$\tdeps$ is not unusually small, this error leads
to a small error in $\tilde\n(\br)$.
But if $\tdeps$ is small, even a small error in 
$v\s(\br)$ can produce a large change in the density, and self-consistency
only increases this effect.  
Thus small $\tdeps$ suggests a large density-error, and the calculation
should be checked.  This is done by inserting an accurate density
in the approximate functional.  If the energy changes significantly, the energy should be substantially more accurate on the exact density.

\begin{figure*}[htb]
\begin{center}
\includegraphics[height=1.5in,width=5in]{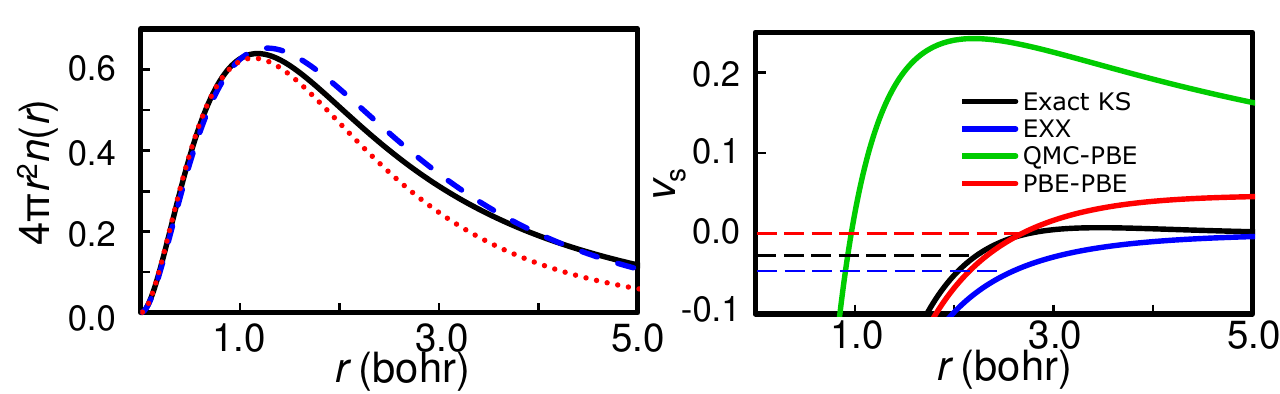}
\caption{Exact\cite{UG94} and approximate densities and KS potentials of H$^-$. The dashed horizontal lines are eigenvalues of 1s orbital with PBE and EXX in a.u..}
\label{Hminus}
\end{center}
~
\vskip -1cm
\end{figure*}
To illustrate this effect in its strongest form, we calculate the energy
of H$^-$.  This anion has two electrons, just like He, but it is long known\cite{SRZ77}
that a standard DFT calculation, in the infinite basis-set limit, cannot bind
two electrons.  In fact, a fraction of an electron is lost to the continuum.
To fully converge such a calculation, we set the occupation of the 1s orbital to, e.g., 1.5, and find a converged solution.  We then slowly increase the occupation until the HOMO level hits exactly zero.  This is then the lowest-energy self-consistent PBE solution.  Its density is very poor (see {\bf Figure \ref{Hminus}}) as it is missing 0.37 electrons\footnote{It is fortunate for approximate DFT that no atomic dianions are bound. To lose one electron may be regarded as a misfortune.  To lose two looks like carelessness.}.  The error in its energy is the same magnitude
as of He (see Table I), but now it is too negative.   On the other hand, the HF
density binds 2 electrons with a negative HOMO, but its energy is very poor.
Finally, the green curve in {\bf Figure \ref{Hminus}} is the PBE potential on the
QMC density.  It has a positive HOMO (really a resonance) and, in a limited
basis set, will yield an accurate self-consistent density (but is not converged).

Of course, the value of DFT is in its computational speed, and would be lost if
we had to calculate a highly accurate density by some other method every time we
ran into a density-driven error.  But because extreme density-driven errors are due to the lack of derivative discontinuity in the
energy, which is reflected in the XC potential, a HF density from an orbital-dependent functional, does not suffer from such errors, and is exact for one electron. Thus the HF density is better for such systems, as we show below.  So we name
the method HF-DFT, meaning to use a HF density with DFT energies.
From {\bf Table \ref{tab1}}, we see that, evaluating PBE on the QMC density of H$^-$, yields an incredibly accurate answer.  HF-DFT also substantially improves over self-consistent DFT but, because this case is so severe, further improvement is gained from the QMC density.

Technically, it is not so easy to precisely perform a KS
calculation on a HF density, as one must find the KS potential
by a process of inversion, which can be complicated and difficult to converge.
A simple workaround is to approximate the HF-DFT energy as
\ben
E^{\rm HF-DFT} \approx E\HF + (\tilde E\xc[\n\HF]-E\x\HF).
\een
Because of the variational principle, this is accurate to second-order in the
density difference, which turns out to be good enough.   On a website\footnote{http://tccl.yonsei.ac.kr/mediawiki/index.php/DC-DFT}
one can find instructions that perform this procedure for several standard codes.
The basic trick is to take the output density of a converged HF calculation,
and feed it into a DFT cycyle, but set the number of iterations to zero or one depending on the code.

\ssec{History of HF-DFT}

The use of HF densities in DFT calculations has a long history.  Even before the
mid 90's, it was common practice to test approximate density
functionals on HF densities\cite{B88,SSP86}.
When DFT was first becoming
popular for routine calculations on main group elements, the initial calculations
were performed on HF densities, in order 
to compare ``apples-to-apples"\cite{GJPF92,OB94}. 
 Pioneering work even noted that, in difficult cases, HF densities somehow yielded better
results than self-consistent results\cite{S92}.  More recently, the
improvement in barrier heights of transition states has been repeatedly
observered\cite{JS08,VPB12}.

But what was previously missing was a general explanation for these better results, and a way to predict when HF-DFT would be better than self-consistent DFT.   In fact, for normal systems, HF-DFT is often slightly worse, as we saw for the He atom, and in many ways, the HF density is less accurate than the self-consistent DFT density\cite{GMB16}. 
  Moreover, the theory given in Section \ref{ksdft} is entirely general, applying to {\em any} 
approximate DFT calculation, not just those with semilocal functionals.  Thus our method explains how and when HF-DFT  is a useful idea.

\ssec{Electron affinities}

The origin of the current theory lies in the calculation of electron affinities
with DFT.  For many years, the Schaefer group successfully calculated
electron affinities within DFT\cite{RTS02}. By using the same basis
for both the anion and the neutral species, finding the DFT energy difference,
and increasing the basis set until the answer stopped changing.  This worked
in many cases, especially those of biological interest\cite{DS09, GXS10, GXS10b, CGCS10, KS10}.  A slight flaw was that the HOMO of the anion would be positive (see {\bf Figure \ref{Hminus}}), which meant these calculations were unconverged\cite{RT97}.

The answer to this apparent conundrum is given by the green curve of {\bf Figure \ref{Hminus}}.  Although the HOMO is technically a resonance, the width of the barrier holding the electron in is so wide that any standard basis functions will not detect the lower-energy state outside the barrier.  Hence the reasonable performance and apparent convergence of electron affinities.  However, the truly converged result is the one
mising a fraction of an electron, which has a terrible energy({\bf Table \ref{tab1}}).

But this also suggested an alternative, more satisfying approach.  Since the
problem is with the self-consistency of the density, if a more accurate
density was available (in this case, a bound one), it should also work.
Thus HF-DFT was used, and found to give comparable (or better) results
for atoms and small molecules.  In fact, using this method, electron affinities
are typically twice as good as ionization potentials with approximate
DFT\cite{LB10,LFB10}.   It should be used for all anionic DFT calculations in future.
It was quite surprising that no-one seemed to have applied this logic
to the anion problem in DFT before.

The explanations in the papers addressing electron affinities
are given in the language of self-interaction error\cite{LB10,LFB10,KSB11}.  This was later generalized
to the general energy-error analysis discussed here, when it was discovered
that DFT calculations on radicals can also be improved with HF-DFT, even though no species are charged\cite{KSB13,KSB14}.

We use HF densities because they are computationally accesible
for molecular systems.  We should use the exact density, but
often HF densities are sufficiently good that any
remaining density-driven error is much smaller than the functional-error. But
HF densities are not always good enough, or even a good choice.  For example,
for H$^-$, the HF-PBE energy is -0.521 Hartree, which is substantially different from the QMC-PBE energy(-0.527 Hartree). 
Another typical failing is when the HF calculation suffers from substantial
spin contamination.  Then the HF density is certainly not accurate enough, and a more advanced method must be used.  Finally,
we mention that for solids, especially metals, HF calculations can be very
expensive and problematic, so in this case, some other method would be better
for calculating an accurate density (for an abnormal system).

Affinities involved in the successive fluorination of ethylene are
afflicted by positive HOMO's and the standard
basis set treatment fails\cite{PDT10}.
For the most extreme example, see also Refs \cite{MUMG14,GB15}.
Using a reasonable basis set
is often used to coax an electron affinity from a standard
functional when evaluating on a data base involving anions\cite{CPR10,HSXR13}.
Many authors emphasize the importance of the basis for
DFT calculations of electron affinity\cite{CHKK15,CFDH15}, and some have
explored the
difficulties in extracting electron affinities\cite{TDGT14}.
The relation between derivative discontinuities, delocalization
error and positive HOMO values is extensively explored in 
Ref. \cite{PTHT15}.
The importance of exact exchange has also been noticed
for genuinely meta-stable anions\cite{FDAB14}, where the
HOMO {\em is} positive.

\ssec{Binding curves}
\label{e:binding}

Our next abnormality is a well-known failing of standard
DFT approximations\cite{RPC06}.
KS-DFT calculations of molecular dissociation energies ($E_{b}$) are
usefully accurate with GGA's, and more so with hybrid functionals.
These  errors are often about 0.1 eV/bond\cite{ES99}, but
are found by subtracting
the calculated molecular energy at its minimum from the
sum of calculated atomic energies.  

However, things look very different if one calculates a binding
energy curve by simply plotting the molecular energy as a function
of atomic separation $R$.
This is because,
if one simply increases
the bond lengths to very large values, the fragments fail to
dissociate into neutral atoms.  
Incorrect dissociation occurs whenever the approximate HOMO of one atom is below
the LUMO of the other\cite{RPC06}, which guarantees a vanishing $\tdeps$
when the bond is greatly stretched.   
With standard functionals, this happens for more than half of all heteronuclear
diatomic pairs.
The exact $v\xc(\br)$ contains
a step between the atoms which
is missed by standard approximations.  Since the step is often dominated by 
the exchange term, effectively only a fraction of this step is contained
in a hybrid calculation.  In the very stretched case, this effect can
also be explained in terms of the inability of the approximations
to reproduce the derivative discontinuity in the energy.

\begin{figure}[htb]
\begin{center}
\includegraphics[height=1.5in,width=3.5 in]{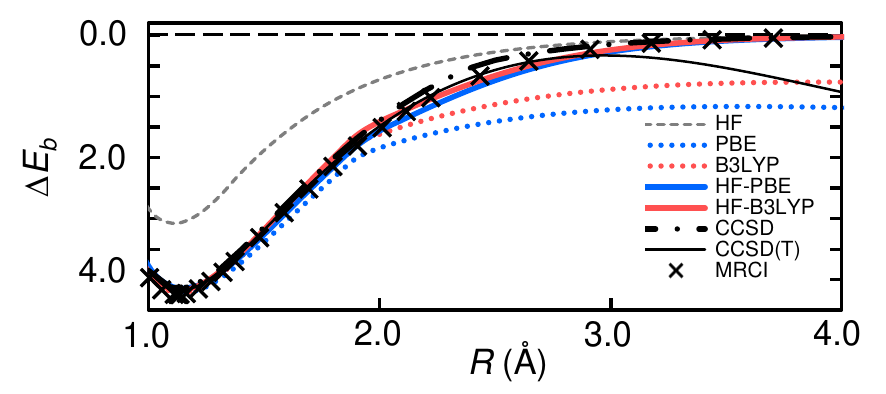}
\vskip -0.5 cm
\caption{Binding energy curves of CH$^+$ with various methods. The cross mark is MRCI results from Ref. \cite{BSM14}. Energies are in eV.} 
\label{CHp}
\end{center}
~
\vskip -1 cm
\end{figure}

A recent paper\cite{KPSS15} describes how to perform HF-DFT
calculations that both overcome the dissociation limit problem, and yield
accurate binding energy curves out to much larger separations than was
previously possible.  A beautiful example is posed by a molecule that is 
very challenging to theory, CH$^+$ (but of perhaps limited interest experimentally).
All DFT methods perform satisfactorily near the bond minimum, yielding
accurate atomization energies when subtracted from the corresponding atomic
calculations of C$^+$ and H.  They can be compared with the `gold standard'
of ab initio quantum chemistry, CCSD(T).  However, as the bond is stretched,
it becomes multi-reference in character, and even CCSD(T) fails badly.  The
perturbative treatment of triple excitations fails as the gap shrinks to zero.
CCSD behaves better, but only a multi-reference configuration interaction (MRCI) calculation
yields an accurate curve.
Self-consistent DFT yields incorrect dissociation limits and, even worse,
deviates from the accurate curve at only 2\,\AA . 
However, {\bf Figure \ref{CHp}} shows HF-DFT works extremely well here, closely following
the accurate curve out much further, as well as producing the
correct dissociation products, for most approximate functionals.

\ssec{Potential energy surfaces of radical and charged complexes}

There are many branches of chemistry in which either radicals or anions,
dissolved in water, are vitally important.  To perform ab initio molecular dynamics simulations of such systems, KS-DFT calculations must yield accurate
potential energy surfaces for the complexes. DFT calculations with standard functionals often yield incorrect global minima with fictious hemi-bonds appearing, in which the additional electron localizes halfway between two species. This is blamed on self-interaction error.
HF-DFT cures all these problems, making potential energy surfaces
highly accurate with any of the popular functionals\cite{KSB14}.

\ssec{Applications of energy-error analysis and other approaches}

There are already many applications in the literature where the
energy-error analysis has been applied to calculations with abnormal
standard approximations.
As the length of a long-chain hydrocarbon grows, the ionization potential collapses to the KS HOMO level with standard approximations, due to the incorrect delocalization of the hole over the entire molecule\cite{WVIJ15}.  This effect should not be present in HF-DFT, but that has yet to be tested.
Gaps have been analyzed to see if a strong density-driven error is responsible
for poor performance for RNA backbones\cite{KMGH15}.
The energy-error analysis has also been used in analyzing errors in 3-body DFT energies\cite{G14}.
The delocalization error has been implicated in difficulties calculating
alkylcobalamins\cite{GNPM13}, where HF-DFT might be very helpful.
It has also been useful in analyzing intercage electron transfer in Robin-Day type molecules\cite{WLZL12}, and a 
large density-driven error has been found in the Kevan model of a solvated electron\cite{JOD13}.
IR spectra of small anionic water clusters have been shown to be problematic
with fixed basis DFT calculations, and fixed by MP2\cite{GDTJ15}.  But HF-DFT
has not been tried, and should be better than MP2.

There are also cases where HF-DFT has definitely improved results.
HF-DFT has been used (successfully) to deal with anion, dianion, and radical Fullerene
oligomers\cite{SSSZ14}.
In diene isomerization, the energy-error analysis has been performed, with a strong suggestion
that inconsistency improves energetics\cite{WPAS15}.
Magnetic exchange couplings can be greatly improved by inconsistent calculations\cite{PP12},
and might also be relevant to organic molecules\cite{KCL13}.
It is useful even in estimating metabolic reaction energies\cite{JRDS14}.

Not all suspected density-driven errors turn out to be so, and in those cases,
HF-DFT does not work.
For adhesive energies of hydrogen molecular chains, HF-DFT only slightly improves
over DFT\cite{SX15}, presumably because this is not a density-driven error, as all units in the chain are identical just as in {\bf Figure \ref{f1}}.
We have explored HF-DFT as a cure for self-interaction error in anion-$\pi$
complexes\cite{MCRS15}, but found it not to be density-driven.

There are countless other approaches to fixing the problems of abnormal
calculations in the DFT literature.
Range-separated functionals have been shown to cure delocalization
errors of standard functionals in Michael-type reactions\cite{SAR13}, 
but HF-DFT
should also work, while bypassing the need to introduce a system-dependent
parameter.
Other authors have suggested constraining the potentials of DFT
calculations with the correct asymptotic forms\cite{GL12}, which
is another approach ripe for energy-error analysis.
Other alternatives include using Koopman's condition\cite{DFPP13}, or
the use of a model for the exchange hole that avoids the delocalization 
effect on barrier heights\cite{JCDD15}.
The beauty of HF-DFT is that it bypasses the need to find a better
potential or do a more expensive calculation.  It is possibly the most
pragmatic approach to these difficulties, and readily available to any
user.

Of course, more sophisticated (and usually more expensive) calculations
such as RPA usually do not suffer from the specific errors
made by standard approximations\cite{EBF12}.  But many such methods
suffer from acute orbital-dependence:  significantly different
energies are found by using different non-self-consistent orbitals, and
self-consistent calculations are often hideously expensive, both in terms
of computational time and coding, without providing improved results.
These situations are ripe for energy-error analysis.

In fact, many applications of hybrid functionals face a Procrustean dilemma.  
The small value of $a$ is needed to yield accurate energetics\cite{Bb93,PEB96},
but a value closer to 100\% is needed to generate accurate potentials and response
properties (as in {\bf Figure \ref{He}}).  
The use of a local hybrid\cite{BCL98} should overcome the dilemma posed
by global hybrids in this regard\cite{J14}.
Abnormal systems make this problem acute.
But HF-DFT sidesteps the issue, by using a better density without studying the potential.
An ensemble generalization is one of many other approaches to this problem\cite{KSKK15}.

\ssec{Limitations of HF-DFT}

The classic examples\cite{MCY08} of failures of popular DFT approximations
are the binding energy curves of H$_2$ and H$_2^+$.
These two
prototypes illustrate starkly the failures as bonds are stretched, and these
effects happen for most bonds.  Unfortunately, HF-DFT does {\em not} help
here, because of the left-right symmetry in both cases.  Both these errors
are functional-driven, i.e., replacing the self-consistent density with
the exact density makes little difference.  
In {\bf Figure \ref{f1}}), the dashed lines are on the exact (HF) density,
and are very similar to the self-consistent solid lines.

\begin{figure*}[htb]
\begin{center}
\includegraphics[height=1.5in,width=5in]{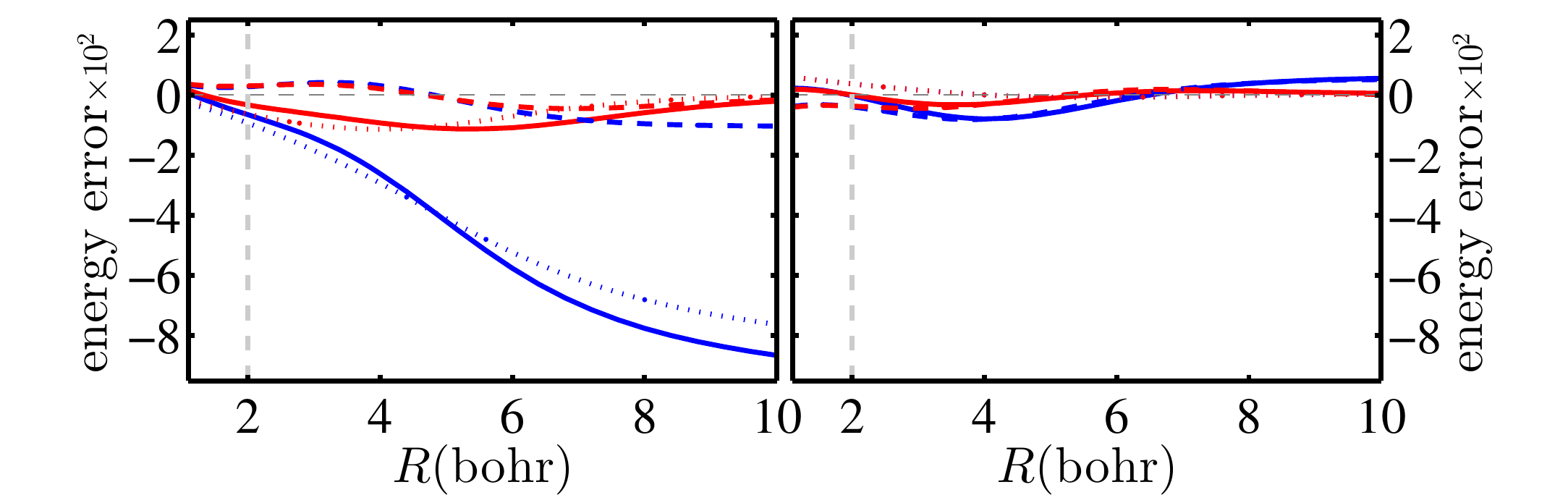}
\caption{Decomposition of the partition energy-error $\Delta E_p$ (left) and total fragment energy-error $\Delta E\f$ (right) for PBE (blue) and
OA (red).  Dotted curves functional-driven, dashed density-driven. Energies are in eV.}
\label{f5}
\end{center}~
\vskip -1cm
\end{figure*}

\sec{PDFT and energy-error analysis}

\ssec{Interpretation of PDFT energies}

Separating functional and density-driven errors can also illuminate the results of embedding calculations \cite{GBMM12} and clarify the role of self-consistency in S-DFT calculations \cite{WS13}. Now we apply the energy-error analysis to a PDFT calculation
in which we know the exact XC functional,
but approximate $\tilde E_p=0$.  Then we trivially find our energy
as the sum of isolated fragments with corresponding fragment densities.
Our energy-error is simply 
\ben
\Delta E = E\f^{(\infty)} - E = -E\d, ~~ \Delta E_F = -E_p, ~~ \Delta E_D = E\rl,
\een
i.e., we can interpret the partition energy as (minus) the functional error
of such a calculation, and the relaxation energy as the density-driven error.
We then say that bonds are {\em normal} when $|E\rl|<<|E\d|$. 
Abnormal bonds are those in which the distortion of the fragment densities relative
to the corresponding atomic densities is sufficiently large to make the
relaxation energy comparable to the dissociation energy.
This definition is precise and unambiguous, and does not depend on any XC approximation.

\ssec{Energy-error analysis within PDFT}

As our last example, we apply the energy-error analysis within an approximate PDFT
calculation.  We use
the OA of Equation \ref{e:OA} on PBE\cite{NW15}.
We write
\ben
\Delta E_p = \Delta E_{p,F} + \Delta E_{p,D}, ~~~ \Delta E\f= \Delta E_{{\rm frag}, F}+\Delta E_{{\rm frag},D}
\een
We plot these in {\bf Figure \ref{f5}}.  
The blue in the left panel of {\bf Figure \ref{f5}} shows that the large error in $E_p$ is functional driven,
as expected.  Even when largely fixed by the overlap approximation, for moderate
bond lengths, this is still functional driven.  But for $R > 5$, the density-driven
error comes to dominate even the OA result, suggesting it can be improved by using the
HF density (just as the heteronuclear bonds of Section \ref{e:binding}).
On the other hand, the fragment errors of the right panel of {\bf Figure \ref{f5}} are much smaller overall.
Moreover, for $R > 4$, these errors are density-driven and so can be reduced with HF-DFT.
For $R < 4$, the density-driven component remains comparable to the functional-driven
piece, which is the same for both PBE and OA.  This strongly suggests that, at least for
H$_2^+$, HF-DFT can reduce the fragment errors, once the principal partition energy-error
has been tackled within PDFT.

\sec{Conclusion}

Emerson\cite{emerson} was clearly referring to DFT and PDFT when he wrote that
{\em a foolish consistency is the hobgoblin of little minds}.
Previously, anyone questioning whether DFT calculations should be self-consistent would be regarded as showing signs of triviality.   We hope to have convinced the readers, possibly for the first time in their lives, of the vital Importance of Being Consistent (when not foolish).


\acknowledgments
The authors are not aware of any affiliations, memberships, funding, or financial holdings that might be perceived as affecting the objectivity of this review. 
A.W., J.N., and K.J. acknowledge support from the Office of Basic Energy Sciences, U.S. Department of Basic Energy Sciences, U.S. Department of Energy, under Grant No. DE-FG02-10ER16191. A.W. also acknowledges support from the U.S. National Science Foundation CAREER program under Grant No. CHE-1149968, and from the Camille Dreyfus Teacher-Scholar Awards Program. M.-C.K. and E.S. acknowledge this work was supported by grants from the National Research Foundation (2014R1A1A3049671) and the Ministry of Trade, Industry \& Energy(MOTIE, Korea) under Industrial Technology Innovation Program. No. 10062161, and the Yonsei Future-Leading Research Initiative. K.B. acknowledges DOE grant no. DE-FG02-08ER46496. O. Wilde is thanked for help with the title and footnotes.


\begin{widetext}

\printbibliography

\end{widetext}

\end{document}